\documentclass{article}
\usepackage{spconf,amsmath,graphicx}
\usepackage{bm}
\usepackage{multirow}
\usepackage{float}
\usepackage{makecell}
\usepackage{tablefootnote}
\usepackage{booktabs}
\usepackage{caption} 
\usepackage{hyperref} 
\usepackage{algorithm}
\usepackage{algorithmic}
\usepackage{pifont} 
\usepackage{amssymb}
\usepackage{marvosym}

\usepackage{color}
\definecolor{applegreen}{rgb}{0.55, 0.71, 0.0}
\definecolor{alizarin}{rgb}{0.82, 0.1, 0.26}

\title{CIF-T: A Novel CIF-based Transducer Architecture for \\ Automatic Speech Recognition}
\name{
    \begin{tabular}{c}
       \large ~Tian-Hao Zhang$^{1, 2, *}$,~Dinghao Zhou$^{1, *}$\thanks{* Equal contribution.}, ~Guiping Zhong$^{1}$, ~Jiaming Zhou$^{1, 3}$, ~Baoxiang Li$^{1,} \textsuperscript{\Letter}$\thanks{\textsuperscript{\Letter} Corresponding author.}
   \end{tabular}
    \vspace{-5pt}
}
\address{
\large
    $^1$  SenseTime Research \\
    $^2$ University of Science and Technology Beijing \\
    $^3$ Nankai University
\vspace{-10pt}
}

\begin{document}
\ninept
\maketitle

\begin{abstract}
RNN-T models are widely used in ASR, which rely on the RNN-T loss to achieve length alignment between input audio and target sequence. However, the implementation complexity and the alignment-based optimization target of RNN-T loss lead to computational redundancy and a reduced role for predictor network, respectively. In this paper, we propose a novel model named CIF-Transducer (CIF-T) which incorporates the Continuous Integrate-and-Fire (CIF) mechanism with the RNN-T model to achieve efficient alignment. In this way, the RNN-T loss is abandoned, thus bringing a computational reduction and allowing the predictor network a more significant role. We also introduce Funnel-CIF, Context Blocks, Unified Gating and Bilinear Pooling joint network, and auxiliary training strategy to further improve performance. Experiments on the 178-hour AISHELL-1 and 10000-hour WenetSpeech datasets show that CIF-T achieves state-of-the-art results with lower computational overhead compared to RNN-T models.

\end{abstract}

\vspace{-3mm}
\section{Introduction}
Recurrent neural network transducer (RNN-T) models \cite{rnn1, rnn2, rnn3, rnn4} have gained significant attention because of their natural streaming capability and superior performance in ASR tasks. RNN-T is initially proposed to address the conditional independence assumption of CTC models by introducing a predictor network that serves as a language model (LM). During RNN-T training, \textit{blank} symbols are added with RNN-T loss to facilitate the learning of alignments between acoustic and semantic features, making RNN-T models particularly suitable for frame-synchronous decoding. However, RNN-T needs to consider all feasible decoding paths, as illustrated in Fig. 1, which requires the probability distribution of all symbols in the utterance at each time step (usually a 4-D tensor) \cite{DBLP:conf/interspeech/KuangGKLLYP22}. This results in a high demand for training resources, which leads to a much longer training times. Similarly, excessive computational redundancy causes high prediction delay in the decoding process.

\begin{figure}[t]
  \centering
  \includegraphics[width=\linewidth]{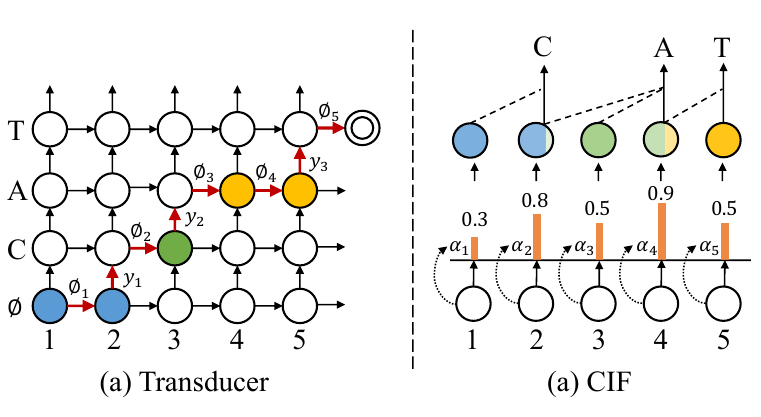}
  \vspace{-6mm}
  \caption{The different aggregation processes of acoustic features between RNN-T and CIF. The RNN-T emits special symbols $blank$ for the alignment process, while CIF aggregates the weighted $\alpha$ of acoustic features.}
  \label{fig:speech_production}
  \vspace{-6mm}
\end{figure}

Numerous efforts have been made to decrease the computational redundancy of RNN-T. Li et al. \cite{li2019improving} remove the padding portion of the encoder and predictor network outputs, and use a sentence-by-sentence combination instead of broadcasting. Ref \cite{wang2022accelerating} first predicts the posterior probability of the encoder outputs with CTC \cite{DBLP:conf/icml/GravesFGS06} before feeding them to the joint network, and then removes the frames predicted for the symbol \textit{blank} according to a specific threshold. Considering the extensive vocabulary size, Kuang et al. \cite{DBLP:conf/interspeech/KuangGKLLYP22} propose Pruned RNN-T, which restricts the range of predictor network output at each time step to minimize the computation of RNN-T loss. Other works focus on optimizing the decoding path of RNN-T to decrease the delay in decoding process \cite{DBLP:conf/interspeech/KimLTZS21, DBLP:conf/icassp/YuCLCSHNHGWP21, DBLP:conf/interspeech/SainathPRGS20, DBLP:conf/icassp/InagumaGLLG20}.

\begin{figure*}[t]
  \centering
  \setlength{\abovecaptionskip}{-1.mm}

  \includegraphics[scale=0.42]{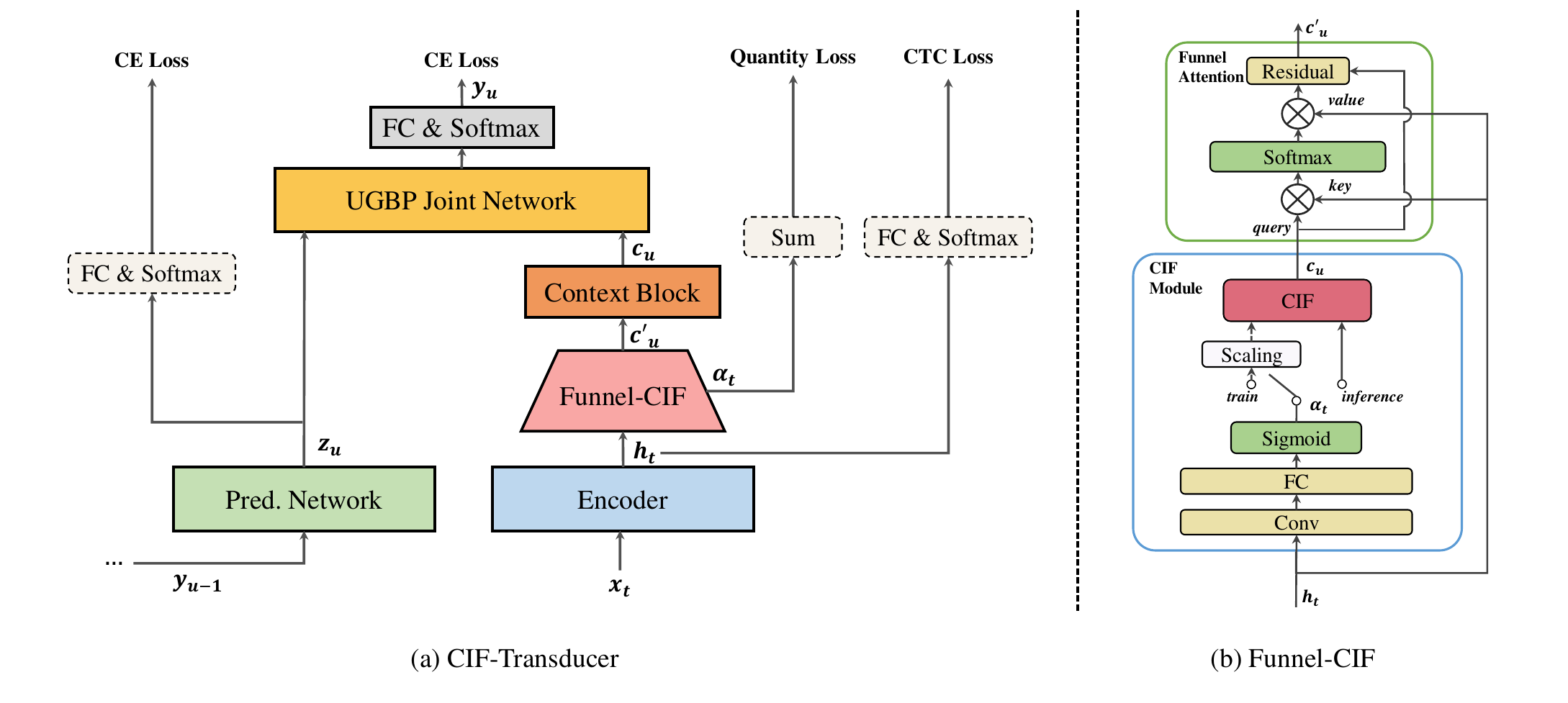}
  
  \caption{The structure of the proposed CIF-Transducer and Funnel-CIF. The dashed boxes in Fig. (a) represent the modules used only for the training process. FC and Conv stand for fully connected layer and convolutional layer, respectively.}
  \label{fig:all_model}
  \vspace{-0.5cm}
\end{figure*}

Although these methods have been successful in decreasing the redundant computation of RNN-T models, they are still the improvement of RNN-T loss. However, the utilize of RNN-T loss to constrain the model gives rise to another significant challenge. The primary optimization target of RNN-T loss is to achieve length alignment between the input audio and the target sequence, which result in an over-reliance on the predictor network to facilitate alignment. This over-reliance come at the expense of sacrificing the essential contextual semantic knowledge required for accurate transcription. As discussed in \cite{DBLP:conf/icassp/ShrivastavaGCZS21}, when substituting the weights of the predictor network with randomized values, the RNN-T model quality remains almost the same as the fully trainable baselines. This constrains the capacity of RNN-T for internal language modeling.

The primary issue causing these problems is the difficulty in handling the length mismatch between the input audio and target sequence, which poses a challenge for designing the RNN-T loss and fuse the two modalities effectively in the joint network. To address this challenge, we observe that the \textit{blank} symbols emitted in the RNN-T decoding process are essentially an aggregation of acoustic features, and the successful emission of non-blank characters indicates the availability of sufficient acoustic information for prediction. As illustrated in Fig. \ref{fig:speech_production}(a), the blue acoustic features are aggregated as the \textit{blank} symbols are emitted, and when $\mathbf{h}_1$ and $\mathbf{h}_2$ complete the aggregation, the character $C$ is emitted. The green acoustic feature $\mathbf{h}_3$ is used to emit $A$, and the yellow acoustic features $\mathbf{h}_4$ and $\mathbf{h}_5$ are used to emit $T$ as the aggregation continues. This mechanism is consistent with the recently proposed soft and monotonic alignment method, known as Continuous Integrate-and-Fire (CIF) \cite{DBLP:conf/icassp/Dong020, dong2020comparison}. As depicted in Fig. \ref{fig:speech_production}(b), CIF accurately identifies acoustic boundaries and extracts acoustic features corresponding to each target symbol for prediction by accumulating the weighted $\alpha$ at each time step. The blue, green, and yellow feature blocks indicate the corresponding acoustic features used to predict $C$, $A$, and $T$, respectively. Hence, we replace the RNN-T loss with CIF module, significantly reducing the computational burden in the joint network and directly supervising the posterior probability of each predicted symbol with cross-entropy (CE) loss.

In this paper, we propose a novel architecture for ASR named CIF-Transducer (CIF-T), which incorporates the CIF mechanism with the RNN-T model to achieve efficient alignment. Our approach obviates the requirement for the RNN-T loss, resulting in a substantial reduction in computational redundancy and allowing predictor network to assume a more prominent role in enhancing the predictions accuracy. Due to the monotonic alignment property of CIF, it is seamless to integrate the CIF mechanism with the RNN-T model while maintaining the ability to decode in a streaming manner. During the alignment process utilizing the CIF module, a certain amount of original acoustic information may be lost. In order to mitigate this loss, we propose an extension to the CIF module called Funnel-CIF, which supplements the original information to the CIF outputs. Moreover, we introduce the Context Blocks, Unified Gating and Bilinear-Pooling (UGBP) joint network, and auxiliary training strategy to further enhance the performance. We conduct experiments on the public available 170-hour AISHELL-1 and 10000-hour WenetSpeech datasets, our proposal achieves state-of-the-art results while significantly reducing computational overhead compared to vanilla RNN-T models. Our work is the first to empower CIF-based models to surpass the performance of other auto-regressive models.
\vspace{-4mm}
\section{Related Work}
\vspace{-2mm}
\subsection{RNN Transducer}
\label{rnnt}
The RNN-T model consists of three components, which are encoder, predictor network, and joint network, respectively. Given the acoustic input $\mathbf{X}_\mathrm{T}$ and the corresponding target sequence $\mathbf{Y}_\mathrm{U}$, where $\mathrm{T}$ represents the length of the acoustic input and $\mathrm{U}$ denotes the number of symbols in $\mathbf{Y}$, the output $\mathbf{h}^{\mathrm{joint}}_{t,u}$ of RNN-T is calculated as follows:
\vspace{-2mm}
\begin{align}
    \label{eq1}
    &\mathbf{h}_t=\mathrm{Encoder}(\mathbf{x}_{:t}) \\
    &\mathbf{z}_u=\mathrm{PredictorNetwork}(\mathbf{y}_{:u}) \\
    &\mathbf{h}^{\mathrm{joint}}_{t,u}=\mathrm{JointNetwork}(\mathbf{h}_t, \mathbf{z}_u)
\end{align}
the semantic feature of the $u$-th symbol produced by the predictor network. 
After getting $\mathbf{h}^{\mathrm{joint}}_{t,u}$ via the joint network, the posterior probability of the predicted symbol is:

\vspace{-5mm}
 \begin{gather}
      \label{eq_p}
       \mathrm{P}(\hat{\mathbf{y}_i}|_{\mathbf{y}_{:u-1},\mathbf{x}_{:t}})=\mathrm{Softmax}(\mathbf{W}_{\mathrm{o}}\mathbf{h}^{\mathrm{joint}}_{t,u})
 \end{gather}
 where $\mathbf{W}_{\mathrm{o}}$ is the weight of the classification layer, whose output dimension is $\mathrm{V}$, representing the number of all symbols that containing \textit{blank}. $\hat{\mathbf{y}_i}$ is belong to $\hat{\mathbf{y}}=\{\mathbf{y}_1,\mathbf{y}_2,...,\mathbf{y}_\mathrm{U+T}\}$, which denotes a possible RNN-T decoding path, and $\mathbf{Y}_\mathrm{U}$ can be derived from $\hat{\mathbf{y}}$ by removing all $blank$ symbols in it. During training, we use $\Omega$ to represent all possible decoding paths, the RNN-T loss is given as:
 \vspace{-1mm}
  \begin{gather}
      \label{eq_l}
      \mathcal{L}(\mathbf{Y}_\mathrm{U})=-\sum_{\hat{\mathbf{y}} \in \Omega} \prod_{i=1}^{\mathrm{T+U}}\mathrm{Log} \mathrm{P}(\hat{\mathbf{y}}_i|\mathbf{x}_{:t_i},\mathbf{y}_{:u_i})
 \end{gather}
 where $t_i$ and $u_i$ represent the values of $t$ and $u$ corresponding to $\hat{\mathbf{y}}_i$ in the decoded path $\hat{\mathbf{y}}$, respectively.

\vspace{-2mm}
\subsection{Continuous Integrate-and-Fire}
\label{cif}
    The CIF module is a soft and monotonic alignment mechanism that has recently garnered attention in the speech community and demonstrated success in non-autoregressive attention-based encoder-decoder (AED) ASR models \cite{DBLP:conf/asru/YuLGLYXGHZ21, DBLP:conf/interspeech/GaoZ0Y22, DBLP:conf/asru/HiguchiCFIKLNWW21}. As the CIF module shown in Fig. \ref{fig:all_model}(b), given the encoder output $\mathbf{h}_t$, the weight $\alpha_t$ is obtained by passing it through a convolution layer, a fully connected layer with one output dimension, and sigmoid activation. After that, CIF accumulates $\alpha_t$ forwardly until the weight sum reaches a threshold of $\beta=1.0$, indicating that the acoustic feature boundary has been found. At the boundary weight $\alpha_t$, CIF divides it into two parts, $\alpha_{t1}$ and $\alpha_{t2}$, with $\alpha_t$ equal to $\alpha_{t1}$ and $\alpha_{t2}$. $\alpha_{t1}$ is included in the current acoustic range, while $\alpha_{t2}$ is considered as the starting weight for the next acoustic range. In the determined acoustic range, the acoustic features $\mathbf{h}_t$ are integrated via a weighted summation with the corresponding $\alpha_t$ and firing the integrated acoustic embedding $\hat{\mathbf{c}}_u$ at the boundary. This embedding can then be used in the subsequent to predict the corresponding symbol $\mathbf{y}_u$.

    During training, to ensure that the length of the integrated acoustic embedding generated by the CIF module matches the target text, the scaling strategy $\alpha_u'=\frac{\tilde{\mathrm{S}}}{\sum^\mathrm{U}_{u=1}\alpha_u}$ is adopt. $\tilde{\mathrm{S}}$ represents the length of the target text, and after scaling, the sum of $\alpha_u'$ is equal to the sum of $\tilde{\mathrm{S}}$. Additional, a quantity loss $\mathcal{L}_{Qua}=|\sum^\mathrm{U}_{u=1}\alpha_u-\tilde{\mathrm{S}}|$ is proposed to force the length of the integrated acoustic embedding to be closer to $\tilde{\mathrm{S}}$.
\vspace{-2mm}

\section{CIF-Transducer}
\vspace{-2mm}
The proposed CIF-T retains the three components of vanilla RNN-T as in Section \ref{rnnt}. As shown in Fig. \ref{fig:all_model}(a), the CIF mechanism is used following the encoder to align the length of the acoustic features and semantic features, which are then fused in the joint network. Our experimental results demonstrate that combining the original CIF module with the vanilla RNN-T model already achieves competitive performance with a lower computational cost. However, we strive for even better results with the CIF-T, thus, the Funnel-CIF, Context Blocks, UGBP joint network, and auxiliary training strategy are proposed to improve each component and achieve a superior performance.

\vspace{-3mm}
\subsection{Funnel-CIF and Context Blocks}
As introduced in Section \ref{cif}, the CIF module integrates the acoustic features $\mathbf{H}_\mathrm{T}=\{\mathbf{h}_1,...,\mathbf{h}_t,...\}$ and obtains the integrated acoustic embedding $\mathbf{C}_\mathrm{U}=\{\mathbf{c}_1,...,\mathbf{c}_u,...\}$ by firing, where $\mathrm{U}<\mathrm{T}$. Therefore, the alignment of the CIF module is a dynamic down-sampling process, which is accompanied with the lost of the original acoustic information. Thus, we emplpy Funnel Attention \cite{funnel} after the CIF module to supplement information as the following formulation:
\begin{gather}
    \label{eq2}
    \mathbf{C}'_\mathrm{U}=\mathbf{C}_\mathrm{U}+\mathrm{Attention}(\mathbf{q}=\mathbf{C}_\mathrm{U},\mathbf{kv}=\mathbf{H}_{\mathrm{T}})
\end{gather}
where the query $\mathbf{q}$ for calculating Funnel Attention is the output of the CIF module $\mathbf{C}_\mathrm{U}$, and the vectors key $\mathbf{k}$ and value $\mathbf{v}$ are the original acoustic features $\mathbf{H}_\mathrm{T}$. Then, the reacquired information is supplemented to $\mathbf{C}_\mathrm{U}$ via a residual connection. In this way, more abundant acoustic information is preserved for later joint decoding. Additionally, as mentioned in \cite{DBLP:conf/asru/YuLGLYXGHZ21}, the contextual correlation among the CIF outputs is weak. To address this problem, we employ a series of Conformer layers to act as the Context Blocks, which enhance the contextual dependencies of each token in the CIF outputs. For representation brevity, we still use $\mathbf{C}_\mathrm{U}$ to represent the output of the Context Blocks.
\vspace{-4mm}

\begin{table} 
    \caption{Encoder hyper-parameters for CIF-T of S, M, and L.}
    \centering
    \footnotesize
    \label{hyper}
    \setlength{\tabcolsep}{1.6mm}{
    \begin{tabular}{lccccc}
    \toprule
    Model & Layers & Model Dim & Heads & FFN Dim & Size (M) \\
    \midrule
    CIF-T(S) & 8;2 & 256 & 4 & 2048 & 35 \\
    \midrule
    CIF-T(M) & 15;2 & 256  & 4 & 2048 & 50 \\
    \midrule
    CIF-T(L) & 16;2 & 512 & 8 & 2048 & 130 \\
    \bottomrule
    \end{tabular}}
    \vspace{-0.5cm}
\end{table}

\subsection{Unified Gating and Bilinear Pooling Joint Network}
\label{ugbp}
Many works \cite{DBLP:conf/icassp/ZhangLLSC22,DBLP:conf/icassp/GravesMH13,DBLP:conf/icassp/VarianiRA020,DBLP:conf/icassp/SaonTBK21} suggest that the original fusion using a single linear layer is not effective enough. In this work, we propose the use of the Unified Gating and Bilinear Pooling joint network \cite{DBLP:conf/icassp/ZhangLLSC22} to achieve more effective fusion of acoustic and semantic features. The UGBP first performs dynamic selective fusion of the two modality features in the channel dimension with gating. After that, a low-rank approximation of bilinear pooling is used for a more powerful feature fusion. The procedures can be written as:
\vspace{-4mm}

\begin{align}
    \label{eq2}
    &\mathbf{h}_u^\mathrm{gate}=\mathrm{Gating}(\mathbf{c}_u,\mathbf{z}_u) \\
    &\mathbf{h}_u^\mathrm{bi}=\mathrm{BilinearPooling}(\mathbf{c}_u,\mathbf{h}_u^\mathrm{gate})
\end{align}

\begin{table}[t] 
    \caption{Results on the AISHELL-1 dataset. $\dagger$ denotes the best results in the paper. $\ddagger$ represents the vanilla model without improvements.}
    \vspace{-2mm}
    \centering
    \renewcommand{\arraystretch}{1.0}
    \footnotesize
    \label{aishell}
    \setlength{\tabcolsep}{3.7mm}{
    \begin{tabular}{l cc c}
    \toprule
    \multirow{2}{*}{\textbf{Models}} & \multirow{2}{*}{\textbf{Size (M)}}  & \multicolumn{2}{c}{\textbf{CER (\%)}} \\
    & & \makecell{Dev} & \makecell{Test} \\
    \midrule
    \textbf{Conformer AED} \\     \label{note1}
    \quad ESPnet$\tablefootnote{\label{note1}https://github.com/espnet/espnet/tree/master}$ \cite{espnet} & 46 & 4.3 & 4.6  \\
    \quad Wenet$\tablefootnote{\label{note2}https://github.com/wenet-e2e/wenet/tree/main}$ \cite{wenet} & 47 &  - & 4.6  \\
     \textbf{CIF based}  \\
     \quad Conformer-CIF \cite{DBLP:conf/asru/YuLGLYXGHZ21} & 45 & 4.8 & 5.3  \\
     \quad Conformer-CIF$^{\dagger}$ \cite{DBLP:conf/asru/YuLGLYXGHZ21}  & 55& 4.5 & 4.9 \\
     \quad Paraformer$\tablefootnote{https://modelscope.cn/models/damo/speech\_paraformer\_asr\_nat-zh-cn-16k-aishell1-vocab4234-pytorch}$ \cite{DBLP:conf/interspeech/GaoZ0Y22}  & 46 & 4.7 & 5.1 \\
     \textbf{RNN-T based} \\
     \quad K2$\tablefootnote{\label{note3}https://github.com/k2-fsa/icefall/tree/master}$ \cite{DBLP:conf/interspeech/KuangGKLLYP22} & - &4.8& 5.0 \\
     \quad ESPnet\footref{note1} \cite{espnet}  & 35 & 4.3 & 4.8 \\
     \quad Wenet\footref{note2} \cite{wenet} & 53 & -  & 4.5\\
    \midrule
    \textbf{RNN-T (ours)} \\
    \quad RNN-T$^{\ddagger}$ & 35 & 4.7 & 5.3 \\
    \quad \quad + UGBP & 38 & 4.5 &  5.0  \\
    \midrule
    \textbf{CIF-T (ours)} \\
    \quad CIF-T$^{\ddagger}$(RNN-T$^{\ddagger}$ + CIF) & 35 & 4.6 & 5.3 \\
    \quad CIF-T(S) & 35 & 4.4 & 4.8\\
    \quad CIF-T(M) & 50 & 4.3 & 4.5\\
    \quad CIF-T(L) & 130 & \textbf{4.1} & \textbf{4.3} \\
    \bottomrule
    \end{tabular}}
\vspace{-10mm}
\end{table}

\vspace{-1mm}
Thanks to the usage of the CIF mechanism, we achieve the length-aligned joint network inputs $\mathbf{C}_\mathrm{U}$ and $\mathbf{Z}_\mathrm{U}=\{\mathbf{z}_1,...,\mathbf{z}_u,...\}$ in advance. Thus we can simply sum the two modal features and still maintain the original three-dimension, contrary to the traditional RNN-T, which requires summing the two by broadcasting to obtain a four-dimensional tensor. With this, the computational overhead of fusion is significantly reduced. Finally, the output of UGBP is obtained after shortcut connection and tanh activation.
\vspace{-4mm}

\begin{gather}
    \label{eq2}
    \mathbf{h}_u^\mathrm{joint}=\mathrm{tanh}(\mathbf{h}_u^\mathrm{bi}+ \mathbf{W}_1\mathbf{c}_u+\mathbf{W}_2\mathbf{z}_u)
\end{gather}
where $\mathbf{W}_1$ and $\mathbf{W}_2$ are the weights of fully connected layer for the transformation of $\mathbf{c}_u$ and $\mathbf{z}_u$, respectively. The posterior probability of the predicted symbol is similarly calculated by Eq. \ref{eq_p}. Unlike the traditional utilize of RNN-T loss in Eq. \ref{eq_l}, we use CE loss $\mathcal{L}_\mathrm{Joint}$ to constrain the model.

\vspace{-4mm}
\subsection{Auxiliary Training Strategy}
\vspace{-1mm}
In addition to constraining the CIF-T using the CE loss described in Section \ref{ugbp} , we also employ additional auxiliary training objectives. Specifically, we use the CTC loss $\mathcal{L}_\mathrm{CTC}$ and the quantity loss $\mathcal{L}_\mathrm{Qua}$, as described in Section \ref{cif}, to assist in training the encoder and CIF module. Additionally, we utilize CE loss to constrain the predictor network to improve its understanding of contextual semantic information, which we refer to as $\mathcal{L}_\mathrm{LM}$. Fig. \ref{fig:all_model}(a) illustrates the specific implementation of these auxiliary losses. Consequently, the total loss optimized in training is as follows:
\begin{gather}
    \label{loss}   \mathcal{L}=\mathcal{L}_\mathrm{Joint}+\lambda_1\mathcal{L}_\mathrm{LM}+\lambda_2\mathcal{L}_\mathrm{Qua}+\lambda_3\mathcal{L}_\mathrm{CTC}
\end{gather}
where $\lambda_{1:3}$ is hyper-parameters to balance different losses, and the specific values are set experimentally.

\begin{table}
    \caption{Results on the WenetSpeech dataset.}
    \vspace{-1mm}
    \centering
    \renewcommand{\arraystretch}{1.0}
    \footnotesize
    \label{wenetspeech}
    \setlength{\tabcolsep}{1.5mm}{
    \begin{tabular}{l c ccc}
    \toprule
    \multirow{2}{*}{\textbf{Models}} & \multirow{2}{*}{\textbf{Size (M)}}  & \multicolumn{3}{c}{\textbf{CER (\%)}} \\
    & & \makecell{Dev} & \makecell{Test\_Net} & \makecell{Test\_Meeting} \\
    \midrule
    ESPnet \cite{Wenetspeech} & 117 & 9.70 & 8.90 & 15.90  \\
    Wenet \cite{Wenetspeech} & 123 & 8.88 &  9.70 & 15.59  \\
    Conformer-MoE \cite{DBLP:conf/iscslp/YouFSY22} & 425 & 7.67 & 8.28 & 13.96  \\
    \midrule
    CIF-T (L) & 130 & \textbf{7.81} & \textbf{8.73} & \textbf{14.12} \\
    \bottomrule
    \end{tabular}}
\end{table}

\begin{table}[t]
    \caption{Comparison of the maximum batch size trained with RNN-T and CIF-T on a single GPU.}
    \vspace{-1mm}
    \centering
    \footnotesize
    \label{batchsize}
    \setlength{\tabcolsep}{2.6mm}{
    \begin{tabular}{lcccccc}
    \toprule
    \textbf{Batch Size} & \makecell{\textbf{8}} & \makecell{\textbf{16}} & \makecell{\textbf{32}} & \makecell{\textbf{48}} & \makecell{\textbf{72}} & \makecell{\textbf{96}}\\
    \midrule
    RNN-T (Torchaudio) & \checkmark & \checkmark &  \ding{55}  \\
    CIF-T & \checkmark & \checkmark & \checkmark  & \checkmark & \checkmark &  \ding{55} \\ 
    \bottomrule
    \end{tabular}}
    \vspace{-0.4cm}
\end{table}

\section{Experiments}
\vspace{-2mm}
\subsection{Experimental Setup}
\vspace{-2mm}
We conduct experiments on two publicly available datasets, 170-hour AISHELL-1 \cite{Aishell} and 10000-hour WenetSpeech \cite{Wenetspeech}. For all experiments, we represent input vectors as a sequence of 80-dim log-Mel filter bank and set the frame length and shift to 25 ms and 10 ms respectively. The speed perturbation \cite{sp} and SpecAugment \cite{spec} are used before training on both datasets and the features are normalized using Global CMVN \cite{cmvn}. 

We present three versions of CIF-T models, S, M, and L, which share the same two-layer reduced embedding predictor network \cite{DBLP:conf/interspeech/BotrosSDG0H21} and UGBP joint network with same model dimension of 256. We use conformer with 4 times down-sampling CNN layers as our encoder and the different encoder configurations for the three models are shown in Table \ref{hyper}, noting that the values separated by semicolons in the ``Layers" column represent the layers number of Encoder and Context Blocks, respectively. Our experiments are conducted using NVIDIA A100 GPUs, and we use Adam \cite{DBLP:journals/corr/KingmaB14} optimizer with 25000 warm-up steps for AISHELL-1 dataset and 5000 warm-up steps for WenetSpeech dataset. For decoding, we get the final model by averaging selected top-K epochs with the lowest loss on the validation set. The other configurations are same with the presents in WeNet \cite{wenet}. The hyper-parameters $\lambda_1$,$\lambda_2$, and $\lambda_3$ in Eq. \ref{loss} are respectively set to 1, 1, and 0.3 in all experiments.
\vspace{-2mm}

\subsection{Results}
Table \ref{aishell} shows the character error rate (CER) results on the AISHELL-1 dataset. We use CTC loss and LM loss on the vanilla RNN-T model as our baseline model RNN-T$^{\ddagger}$. When just employing the CIF mechanism to replace the alignment process in the RNN-T$^{\ddagger}$, CIF-T$^{\ddagger}$ achieves a competitive result of 4.6\%/5.3\% on the dev/test set compared to 4.7\%/5.3\% for the RNN-T$^{\ddagger}$ as a baseline. With the improvements proposed in this work, the CIF-T(S) and CIF-T(M) achieve consistent or superior performance compared to the best publicly available Conformer-based AED models, CIF-based models, and RNN-T-based models, with equal or smaller parameters. Furthermore, the CIF-T(L) achieves a state-of-the-art result of 4.1\%/4.3\% on the dev/test set. We also evaluate the results of applying UGBP on RNN-T$^{\ddagger}$, and despite the better fusion method improved performance, the results still fall behind our CIF-T models.
The CER results on the WenetSpeech are presented in Table \ref{wenetspeech}. CIF-T achieves a significantly lower CER results than ESPnet and Wenet with essentially the same parameters, while achieving a competitive results compared to Conformer-MoE which has more than three times parameters.
\begin{table}[t]
    \caption{Ablation study of the UGBP joint network, Funnel-CIF, and Context Blocks for CIF-T.}
    \centering
    \footnotesize
    \label{abla}
    \setlength{\tabcolsep}{4mm}{
    \begin{tabular}{lc}
    \toprule
   \textbf{Models} & \textbf{CER}(\%) \\
    \midrule
    CIF-T$^{\ddagger}$ & 5.3 \\ 
    \quad + UGBP & 5.2  \\
    \quad \quad + Funnel-CIF& 5.0\\
    \quad \quad \quad + Context Blocks & 4.8\\
    \bottomrule
    \end{tabular}}
\end{table}

\begin{table}[t] 
    \caption{Influence of re-initializing the predictor network of fully trainable RNN-T and CIF-T.}
    \centering
    \footnotesize
    \label{pred}
    \renewcommand{\arraystretch}{0.9}
    \setlength{\tabcolsep}{4mm}{
    \begin{tabular}{lcl}
    \toprule
    \textbf{Models} & \textbf{Re-Init Pred.}  & \textbf{CER}(\%)\\
    \midrule
    RNN-T  & \ding{55} & \quad 5.0  \\ 
    RNN-T & \checkmark & \quad 5.4 (+0.4) \\ 
    \midrule
    CIF-T  & \ding{55} & \quad 4.8  \\ 
    CIF-T & \checkmark & \quad 6.7 (+1.9)\\ 
    \bottomrule
    \end{tabular}}
    \vspace{-4mm}
\end{table}

Table \ref{batchsize} provides an evaluation of the computational resource consumption of CIF-T and RNN-T models by comparing their respective maximum batch sizes on a single 40G NVIDIA A100 GPU. We utilize the Torchaudio \cite{DBLP:conf/icassp/YangHNACPPGGYLH22} RNN-T loss to train the RNN-T models, and ensure that both models have the same number of parameters. It can be observed that CIF-T can handle a batch size of 72, but exceeds the memory limit when the batch size is increased to 96. In contrast, RNN-T cannot train with a batch size of 32, indicating a larger bottleneck in computational overhead compared to proposed CIF-T.

\vspace{-3mm}
\subsection{Ablation Study and Analysis}
\vspace{-1mm}
Table \ref{abla} shows the performance benefits brought by progressively adding the proposed improvement modules. Experimentally proving that the proposed UGBP, Funnel-CIF and Context Blocks are effective in improving the model performance.

Furthermore, we investigate the role of the predictor network for RNN-T and CIF-T, both models use auxiliary training strategy and UGBP to fuse acoustic and semantic features. As shown in Table \ref{pred}, if we re-initialize the predictor network of the fully trainable RNN-T model randomly, the CER on the test set increases from 5.0\% to 5.4\%, which is only 0.4 percentage points higher, and indicating that the predictor network does not play an indispensable role for correct prediction. Conversely, for the CIF-T model, re-initializing the predictor network led to a substantial increase in CER from 4.8\% to 6.7\%, a difference of 1.9 percentage points. These results suggest that directly constraining the prediction probability with CE loss in CIF-T increases the dependence of the model on the semantic information provided by predictor network. Adopting this approach will be conducive to the integration of improvements related to the predictor network.
\vspace{-3mm}

\section{Conclusion}
\vspace{-2mm}
In this work, we propose a novel architecture for ASR, namely CIF-T. By using an efficient CIF mechanism in the RNN-T model instead of RNN-T loss to achieve length alignment of input audio and target sequence, we achieve a reduction in computational overhead and an enhancement role of predictor network. Additionally, we propose Funnel-CIF, UGBP joint network, Context Blocks and auxiliary training strategy to further improve the performance. Experiments on the AISHELL-1 and WenetSpeech datasets demonstrate that our approach achieves the state-of-the-art results. In the future we will take full advantage of the monotonic alignment property of CIF and explore its application on the streaming models.

\footnotesize   
\bibliographystyle{IEEEtran}
\bibliography{mybib}

\end{document}